# Ion-Implantation as Pixel Isolation for Fabrication of Planar Strain-Balanced Antimony-based Superlattice Infrared Photodetectors

Arash Dehzangi*,

Department of Electrical and Computer Engineering, Northwestern University, Evanston, Illinois 60208, USA

## Abstract

*Strained layer superlattice (SLS) material system is a dynamic and relatively new material for infrared detection. Large format, small-pitch and low-cost focal plane arrays (FPAs) with more pixels are in demand for different applications. For the current SLS based FPAs mesa etching are used to define the pixels. For those SLS based FPAs with scaled pixel size making the mesa structures can be challenging due to the need for deep etch, and then passivation process. One of the possible solutions to address this issue is to consider a planar structure and avoiding the mesa-isolation etching or complex surface treatment/ passivation process. In this work, the recent progress on planar SLS infrared photodetectors using ion-implantation for device isolation is reviewed. In this method of fabrication, ion implantation was applied from the top to bombardment the surface for device isolation, like mesa-isolation step in device fabrication.*

Type-II strained-layer superlattice (SLS) based photodetectors are of great interest as a new material system for infrared photodetectors and imaging systems. Recently, significant development has been made in design, structure, and performance of SLS photodetectors as a new alternative for infrared detector material additional the most mature HgCdTe (MCT) material. [1-5] MCT is II-VI semiconductor alloy grown on nearly lattice-matched CdZnTe (CZT) substrate with strong optical absorption, delivering high performance infrared photodetectors with high quantum efficiency (QE), low dark current and high operating temperature (HOT) capability, covering almost all infrared wavelength detection.[3, 4] However, MCT bears some drawbacks such as bulk and surface growth instability, low yields and higher costs.[1, 6]

SLS material on the other hand offers flexible band gap engineering covering a broad range of cutoff wavelengths along with reliable growth uniformity, high yield and easier manufacturability. [7-13] Compared to SLS based infrared detectors, MCT is a front runner delivering superior performance on QE and dark current. However, thanks to the recent progress in design and growth of unipolar barrier infrared photodetector architecture[14-16], SLS is on the path of development for high-performance infrared photodetectors with reduced imaging system size, weight, and power consumption (SWaP) for focal plane array (FPA) cameras.

*Email: arash.dehzangi@northwestern.edu /arashd53@hotmail.com

Almost all the current SLS based FPAs are fabricated by mesa-isolated pixel approach. The deep mesa etch requirement to isolate the pixels is a major limiting factor due to fabrication challenges for small pixel pitch FPAs along with immediate need to passivate the long and exposed mesa sidewalls. Passivation is another challenge need to be addressed efficiently for small pixel pitch FPAs where the larger perimeter/area ratio makes leakage current to play critical contribution in dark current density. Ample efforts have been made in developing applicable passivation methods to circumvent the surface leakage in narrow band gap SLS photodetectors. [17-19]

To address the issue planar structures with buried junction interface and no passivation requirements have been proposed. The planar architecture itself been implemented for variety of materials including MCT. Multiple design and structures have been reported for planar MCT and InGaAs photodetectors[20-22] and for MCT based FPA production. [23, 24] Diffusion and ion implantation are the most applicable method of junction generation for p-n junction generation in planar devices.

Recently, planar structures fabricated by diffusion and ion implantation techniques been proposed for SLS photodetectors, by avoiding the entire mesa-etching step [25-27]. However, given to the specific epitaxial growth condition and the structure of SLS material, doping with diffusion or ion-implantation along with relevant annealing treatments can have dramatic impact on elegant superlattice interfaces, and drastically degrade the material.

In the case of ion-implantation there is an alternative that instead of doping the material, attempt to perform ion-bombardment on epitaxially-grown doped layers to selectively creates highly resistive regions with damage-related deep levels. [28-30] This high resistive area blocks the electrical crosstalk between neighbor detectors and achieve the same function as mesa isolation. In this work, the recent progress on fabrication of mid-wavelength infrared (MWIR) *nBn* and *pBn* heterostructure InAs/InAsSb type II superlattice photodetector based on the ion-implantation isolation approach will be reviewed. Zn and Si were chosen as the ion implants to create the isolation between the diodes for *pBn* and *nBn* structure, respectively. The electrical and optical performance of the devices were compared together along with mesa-etched devices to make a deeper view of the device performance. The SLS InAs/InAs$_{1-x}$Sb$_x$ (Ga free) was selected due to its superior carrier lifetime along with better controllability in epitaxial growth due to the simpler interface structure. [31, 32] The goal is to entertain the possibility of using the ion-implantation approach for planar SLS photodetectors and pave the way towards future development for FPA applications.

The detail about the material and molecular beam epitaxy (SS-MBE) growth for *nBn* and *pBn* structure was reported before [25, 33]. Both structures have the same barrier and absorption MWIR region, only top contacts are different, as p-type (for *pBn*) and n-type (for *nBn*). The thickness of the top layer has been adjusted accordingly to reflect growth condition. For the barrier, the SLS design was chosen to be AlAs$_{0.5}$Sb$_{0.5}$/InAs$_{0.5}$Sb$_{0.5}$ with a deep electron quantum well in conduction band, which has been demonstrated to act as an effective wide-bandgap electron barrier for both *nBn* and *pBn* structures [33-36].

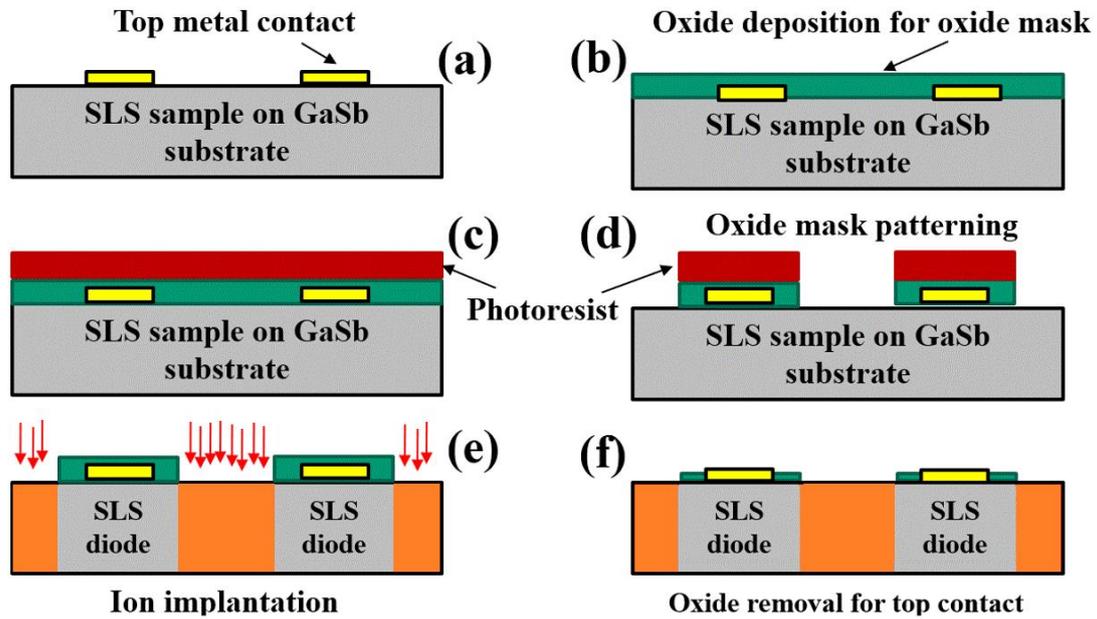

**Figure 1.** Schematic diagram of the process flow for the fabrication planar SLS photodetectors, (a) top metal contact, (b) deposition of the oxide mask, (c,d) patterning via photolithography, (e) implantation for diode isolation (f) top metal contact via.

For the fabrication approach a high quality 800 nm silicon-oxide layer was used as a hard mask to protect the diode area against the implantation. The thickness of the oxide layer was selected based on simulation using the stopping range of ions in matter (SRIM) software. Different steps of fabrication are schematically demonstrated in Figure 1. First step was to deposit top metal contacts (a) on the as-grown surface and then silicon-oxide layer was deposited (b). This oxide layer is designed to act as a hard mask to protect the underneath SLS area as well as the top metal contact. The oxide mask then lithographically patterned on top of the wafers (c, d). Next step was to apply ion implantation on the isolation regions between diodes(e), whose top area were covered by the hard mask. All samples were subjected of annealing for 15 seconds at 300 ºC. Finally top metal contact was opened through the hard mask (f). The shape of the diodes was circular with diameters ranging from 100 μm to 400 μm.

For the implantation process, the dose and acceleration energy required for effective isolation were studied and then optimized using SRIM simulation tool. The penetration depth is function of implanted material nature and implantation energy, where the dose can signify the total number of implanted atoms. Although finding an accurate ion-implantation profile would be hard to estimate, given to the complexities of SLS material and lack of enough empirical data on SLS material ion implantation. Figure 2 illustrates an example of simulation test result from SRIM software for Si implantation distribution in different depths of SLS *pBn* target, performed at two different implantation energies ($I_E$) 100 keV (a) and 190 keV(b). Based on the simulation result and experimental consideration for *pBn* structure ion-implantation energies of 380, 190, and 100 KeV and for *nBn* structure the energies of 300, 190 and 100 KeV were chosen. For both cases and for each implantation energy, three ion-implantation doses of $1.0\times10^{14}$, $5.0\times10^{14}$, and $1.0\times10^{15}$ cm$^{-2}$ were used for a total of 9 permutations. The implantation was performed with a tilt angle of 7° with no cooling (Innovion Corporation). The purpose was to create n-type (silicon) and p-type (zinc) isolation regions on top SLS layer. Both elements are heavy enough ions to generate damage on SLS structure with high carrier removal rates and destructibility to enforce effective isolation between adjacent diodes.[28, 37]

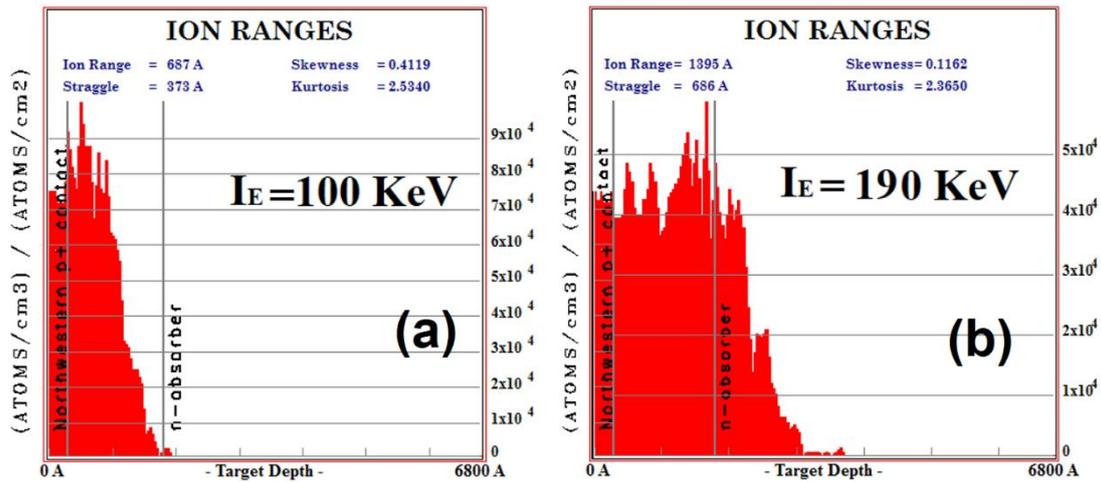

**Figure 2.** Simulation result performed using SRIM software for distribution of the Si ions in different depths of the target for various implantation energies (a) $I_E$ = 100 keV, (b) ) $I_E$ = 190 keV .

To make a meaningful comparison and create a baseline, standard mesa etched photodiodes were also processed from the same MBE grown. The fabrication approach for mesa-isolated diodes was reported elsewhere [38].

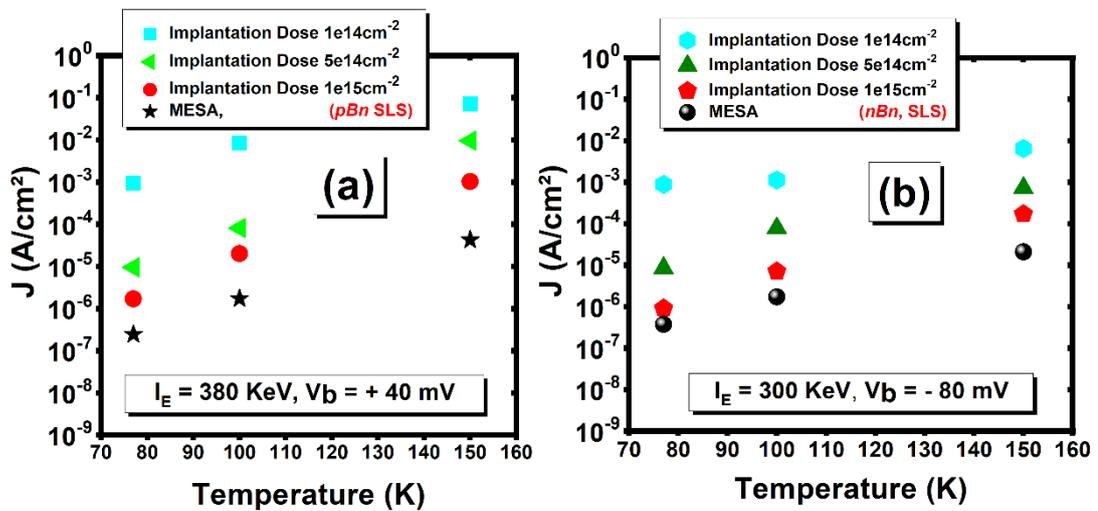

**Figure 3.** a) Dark current values versus temperature for of the planar MWIR SLS diodes (a) *pBn* ($V_b$ = +40 mV, 380 KeV implantation dose) and (b) *nBn* ($V_b$ = -80 mV, 300 KeV implantation dose) fabricated with different implantation doses. All diodes were circular with a 200 µm diameter.

Electrical testing was performed at different temperature for all the SLS samples. The results showed that implantation energy and dose are most critical parameters for electrical performance analysis, where increasing the implantation energy can cause drop in dark current for both *nBn* and *pBn* planar diodes. However, introducing high energy ions to the SLS structure can impose a great risk of damage related increase in dark current such as hopping conduction effect [28, 29]. So unlimited increase of dose or energy of implantation is not recommended and had to be optimized. Given to the implantation technique was used for diode formation, at higher implantation energy each incident ion can reach deeper and make more damage to SLS structure which in turn causing more isolation. Raising the dose can also provide similar impact during implantation and creates effective isolation between adjacent diodes. [28]. The best case is to optimize dose and energy simultaneously to achieve the most effective implantation approach. It was found that the implantation energy of 300 KeV for *nBn* design and 380 KeV for *pBn* design showed the lowest dark current values for each chosen implantation doses. Figure 3a,b demonstrates a comparison of the dark current density versus temperature of planar SLS diodes implanted with different implantation doses for *pBn* at + 40 mV applied bias ($V_b$) (380 KeV implantation energy) (a) and *nBn* ($V_b$ = - 80 mV, 300 KeV implantation energy). The result leads to conclude that the highest dose of $1\times10^{15}$ cm$^{-2}$ to be the choice for optimized condition for each case. Table 1 summarizes the optimized case of implantation for each device. Using the SRIM simulation for these optimized conditions, the estimated values for the depth of ion concentration peak and straggling inside the SLS material are given in Table 1 as well.

Table 1. Optimized implantation parameters for *pBn* and *nBn* design

| Design | Implantation Energy | Implantation Dose | Depth of Ion Concentration Peak | Straggling |
|---|---|---|---|---|
| *pBn* | 300 KeV | $1\times10^{15}$ cm$^{-2}$ | 900 nm | 100 nm |
| *nBn* | 380 KeV | $1\times10^{15}$ cm$^{-2}$ | 1000 nm | 115 nm |

Figure 3 also presents and interesting comparison for the dark current density values for both *pBn* and *nBn* planar and mesa-isolated devices. The result shows that dark current values for mesa-isolated SLS diodes are still lower compared to planar diodes, this issue will be addressed shortly.

The dark current density values versus bias voltage at 77K and 150 K for both optimized implantation condition for MWIR *nBn* and *pBn* planar devices are shown in Figure 4a,b, respectively. All the diodes had 200 μm diameter circle shape. For *nBn* device, the dark current density values at $V_b$ = - 80 mV at 77K and 150 K is $1.23\times10^{-6}$ A/cm$^2$ and $1.42\times10^{-4}$ A/cm$^2$ respectively (Figure 4a). For the *pBn* device at bias voltage of +40 mV, the dark current density values at 77 K and 150K are $5.21\times10^{-6}$ A/cm$^2$ and $2.68\times10^{-3}$ A/cm$^2$, respectively (Figure 4b). In inset of Figure 4 the dark current values for optimized planar SLS photodetectors are compared with traditional mesa-isolated devices for the same size of diodes. As it can be seen at low range of temperature < 100 K, the values for planar devices are in the range of 2-3 (for *nBn*) and 6-8 (for *pBn*) times higher compared to the mesa-isolated devices. The dark current performance is overall better for planar *nBn* device compared to its *pBn* counterpart. The gap between dark current value for planar and mesa-isolated SLS diodes is much worst at higher temperature. At 150 K, the planar devices have more than one order of magnitude higher dark current, which must be addressed especially for high operation temperature (HOT) aspects. This degradation of dark current at high temperature for planar SLS devices can be related to the nature of the defects created by the ion-implantation process and annealing treatment [28]. Further empirical and simulation study of different aspects of ion-implantation for SLS material is strongly recommended. However, the outcome of this research would be useful guideline for future work.

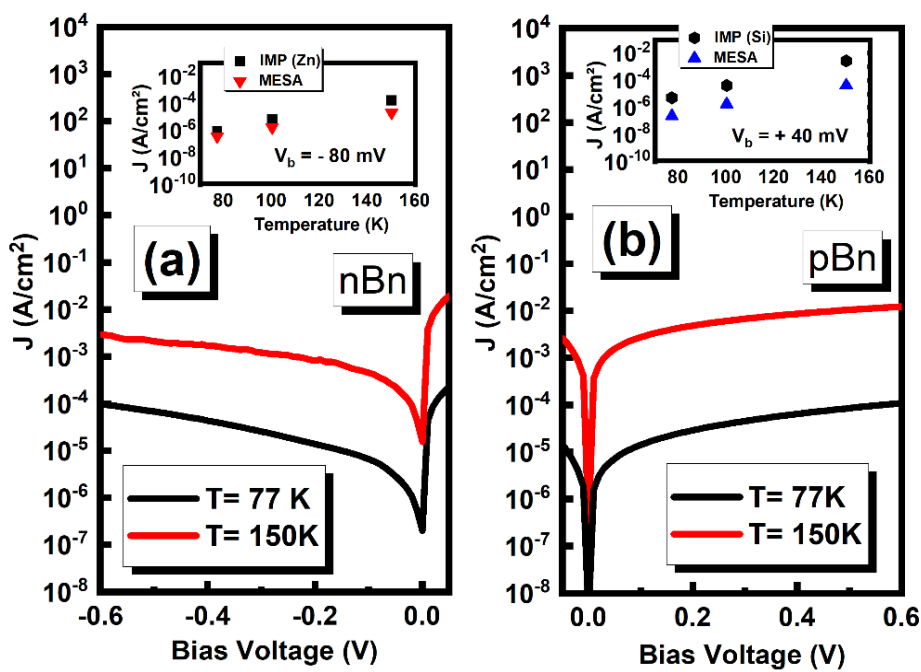

**Figure 4.** a) Dark current density vs. applied bias voltage of the optimized ion-implantation condition at 77K and 150 K for MWIR *nBn* (a) and *pBn* (b) planar devices. The inset of each part compares the dark current density values for planar and mesa etched photodetectors at 77, 100 and 150 K. All diodes were circular with a 200 μm diameter.

The comparison of the optical performance of both MWIR planar devices under optimized implantation at 77 K and 150 k is shown in Figure 5a,b. The calculated saturated QE values of both devices also is compared with the mesa-isolated devices. For optical test, top illumination approach was chosen with no anti–reflection coatings, and a calibrated 1000 °C blackbody source and Fourier transform infrared (FTIR) spectrometer were used. For the *nBn* and *pBn* planar devices QE value is calculated at – 80 mV and +40 mV bias voltage respectively.

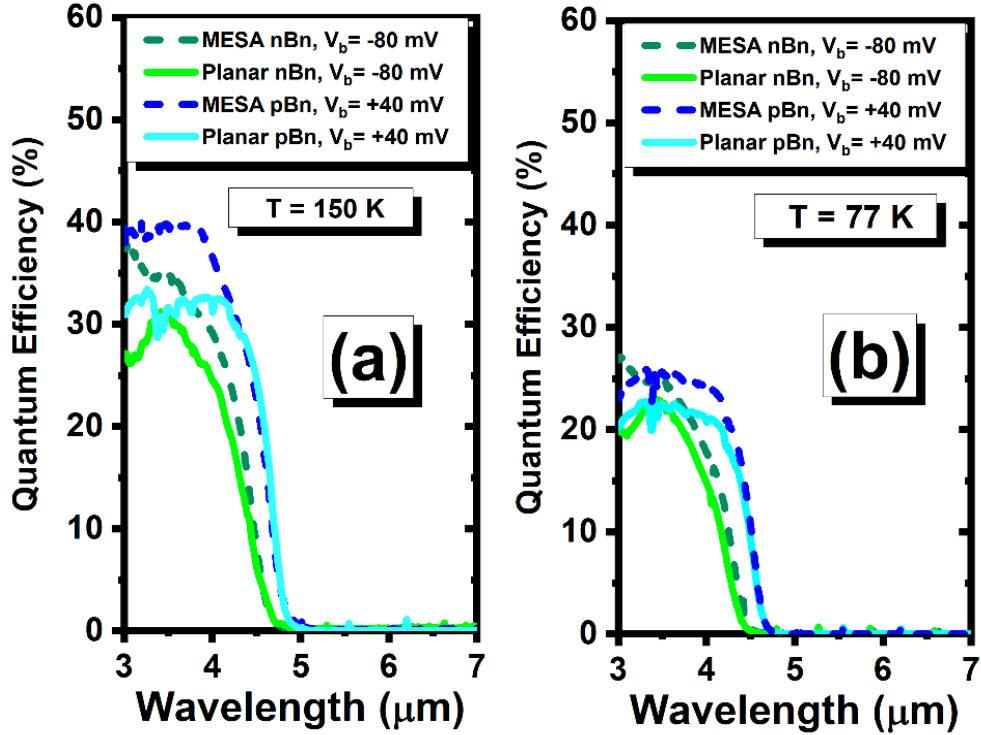

**Figure 5**. Saturated 77 (a) and 150 K (b) QE spectra measured for planar, and mesa etched MWIR SLS photodetectors (diode with 200 μm diameter)

The 50% cut-off wavelength for *nBn* planar device at 77 K, was 4.10 μm with a saturated peak responsivity ($R_i$) of 0.67 A/W at λ = 3.35 μm at – 80 mV applied bias. At 150 K, the saturated peak responsivity was 0.84 A/W at the same wavelength with 50% cut-off wavelength 4.40 μm. For *pBn* device at 77 K and +40 mV bias voltage, the 50% cut-off wavelength of the planar device was 4.40 μm with a peak responsivity of 0.76 A/W at 3.8 μm. At 150 K, the peak responsivity increases to 1.09 A/W. The corresponding saturated QE values at peak responsivity for both planar devices are summarized in Table 2 along with relevant QE values for mesa-isolated devices. For *nBn* device the QE values are lower but close enough to those for the mesa-isolated diodes. For *pBn* design, on the other hand, the gap in QE values is higher compared to mesa isolated devices.

Table 2. Comparison of the QE values at peak responsivity for planar and mesa isolated SLS devices at 77 and 150K.

| Design | Fabrication Approach | QE @ 150K | QE @ 77K |
|---|---|---|---|
| ***nBn*** <br> λ = 3.35 μm | mesa | 36.4% | 25.0% |
| | Planar | 31.5% | 23.5% |
| ***pBn*** <br> λ = 3.80 μm | mesa | 39.2% | 24.4% |
| | Planar | 32.6% | 21.5% |

The reason for the effect is still unknown, but there is a speculation that it might be associated with the optical contribution of the sloped mesa sidewalls (mirror effect) to increase the QE a bit.

The uniformity level of the optical performances of the SLS planar devices was evaluated for optimized implantation parameters to make sure of the proper diode isolation by the implantation technique. To do so, the optical test was performed on different diode sizes at different temperature, and it was confirmed that QE values are not changed across the different range of diode sizes. This is promising for focal plane array applications, which entails scalability of the suggested method. It is worth noting that lower ion implantation energy can cause partial isolation, meaning that the energy is not strong enough to penetrate deeply into the substrate to pass the barrier (i.e., not reaching to absorption region) and generate isolated diodes. This means that the signal could come from the large area of active region under the surface (much larger than the actual size of diode) contributing to the responsivity, which causes inaccurate and large optical response. (e.g., QE> 100%). To confirm this hypothesis further simulation and empirical data is needed, which can be considered for future direction of the study.

A novel approach based on the ion-implantation was implemented for pixel diode isolation for MWIR SLS based heterostructure photodetectors. The different steps in simulation and performing ion-implantation were addressed and the performance of planar nBn and pBn heterostructure InAs/InAsSb SLS photodetectors were compared with each other and with reference mesa-isolated diodes. The planar fabrication method suggests that instead of using ion-implantation for diffusion or doping the substrate, use it to perform ion-bombardment on substrate to selectively form highly resistive regions and isolate the pixels from each other. Both devices revealed similar performance and not very far from their mesa-isolated counterpart devices. Overall, method may have merits to pursue with promising result. There is still some about electrical or optical cross talk between pixels, which might drastically affect the modulation transfer function (MTF) performance of FPAs fabricated by the method, which has to be investigated in further studies. By optimization and developing on structure design of the SLS device and ion-implanting process, it is possible to surpass the mesa-isolated device performance for SLS planar devices.